\documentclass[aps,prl,twocolumn,superscriptaddress]{revtex4}

\usepackage{graphics,amsmath,graphicx}  
\usepackage{pdfpages}
\usepackage[british]{babel} 
\usepackage{bm}
\usepackage{microtype}   

\newcommand{\mc}{\mathcal}
\newcommand{\dg}{\dagger}
\newcommand{\pdg}{{\phantom{\dagger}}}

\newcommand{\Le}{\left}
\newcommand{\Ri}{\right}
\newcommand{\nn}{\nonumber}
\newcommand{\f}{\frac}
\newcommand{\mrm}{\mathrm}

\newcommand{\Yield}{\mathcal{Y}_{>30}}

\newcommand{\fig}[1]{\begin{figure}[!htb]\begin{center}#1\end{center}\end{figure}}

\begin{document}
\title{Impact of quantized vibrations on the efficiency of interfacial charge separation
in photovoltaic devices}

\author{Soumya Bera}
\affiliation{Institut N\'{e}el, CNRS and Universit\'e Grenoble Alpes, F-38042 Grenoble, France}
\author{Nicolas Gheeraert}
\affiliation{Institut N\'{e}el, CNRS and Universit\'e Grenoble Alpes, F-38042 Grenoble, France}
\author{Simone Fratini}
\affiliation{Institut N\'{e}el, CNRS and Universit\'e Grenoble Alpes, F-38042 Grenoble, France}
\author{Sergio Ciuchi} 
\affiliation{Dipartimento di Scienze Fisiche e Chimiche, Universit\`a
  dell'Aquila, CNISM and Istituto Sistemi Complessi CNR, via Vetoio, I-67010
Coppito-L'Aquila, Italy} 
\author{Serge Florens}
\affiliation{Institut N\'{e}el, CNRS and Universit\'e Grenoble Alpes, F-38042 Grenoble, France}

\begin{abstract}
We demonstrate that charge separation at donor-acceptor interfaces is a complex process 
that is controlled by the {\it combined} action of Coulomb binding for electron-hole 
pairs and partial relaxation due to quantized  phonons. A joint electron-vibration 
quantum dynamical study reveals that high energy vibrations sensitively tune the 
charge transfer probability as a function of time and injection energy, due to 
polaron formation.
These results have bearings for the optimization of energy transfer
both in organic and quantum dot photovoltaics, as well as in biological light 
harvesting complexes.
\end{abstract}


\date{\today}
\maketitle

\paragraph{Introduction.}
The dissociation of neutral excitons into separate charge carriers is
a key physical process happening in various light harvesting devices,
such as organic materials~\cite{Bredas09}, quantum dot 
assemblies~\cite{crooker,rozbicki}, and photosynthesis complexes~\cite{scholes2011}.
In the case of organic photovoltaics which is our main focus here, 
an interface is created between a disordered 
polymer (donor), and a fullerene derivative (acceptor).
Because of poor dielectric screening and the narrow bandwidth of the organic acceptor, an electron 
that experiences energetic losses during its transfer at the interface may be
tightly bound to the hole it has left behind. Such an interfacial
exciton displays a binding energy which is typically an order of magnitude 
larger than the thermal energy $k_BT$ at room temperature. 
A highly debated issue~\cite{Lee,Bakulin,Jailaubekov,Gelinas,vandewal}
is therefore whether the exciton created by absorption of 
a photon will eventually relax to the lowest available energy bound state across the 
interface, therefore precluding the coherent migration of the carriers
to the collecting electrode, or if it can separate at sufficiently 
long distances prior to relaxation, enabling photocurrent generation.
These two extreme views are presently difficult to reconcile, but both should certainly 
be at play in the microscopic mechanisms that ultimately control the
efficiency of energy transfer in these systems.

Our aim in this Letter is to study microscopically the role of quantized phonons 
--- i.e. discrete molecular vibrations whose characteristic energy is
larger than $k_BT$ --- in the charge transfer process.
As opposed to low energy baths (due to intermolecular phonons or solvents), which
lead to damping processes and final equilibration on long time scales,  high energy 
intramolecular vibrations correspond to fast time scales in the ten-femtosecond range, 
comparable to the excitonic dissociation process. 
Such quantized modes are strongly coupled to the electronic orbitals, both in 
photosynthetic systems~\cite{Kolli} and in fullerenes, as revealed e.g. by 
{\it ab initio} theoretical calculations ~\cite{Antropov,Faber}, transport \cite{Crespi} 
and photoemission experiments~\cite{Canton}. 
More importantly, their relevance to the charge transfer through organic
interfaces has been pinpointed very recently by ultrafast spectroscopic
techniques \cite{Falke,Song}.  All these observations indicate that high-energy
vibrations should be included in any consistent model for organic photovoltaic devices.
We unveil here that, by coherently dressing the charge carriers, quantized
phonons sensitively control the charge transfer probability, and can be hugely
detrimental to free carrier generation in the presence of interfacial Coulomb
interaction between hole and electron species.
\fig{
  \includegraphics[width=0.9\columnwidth]{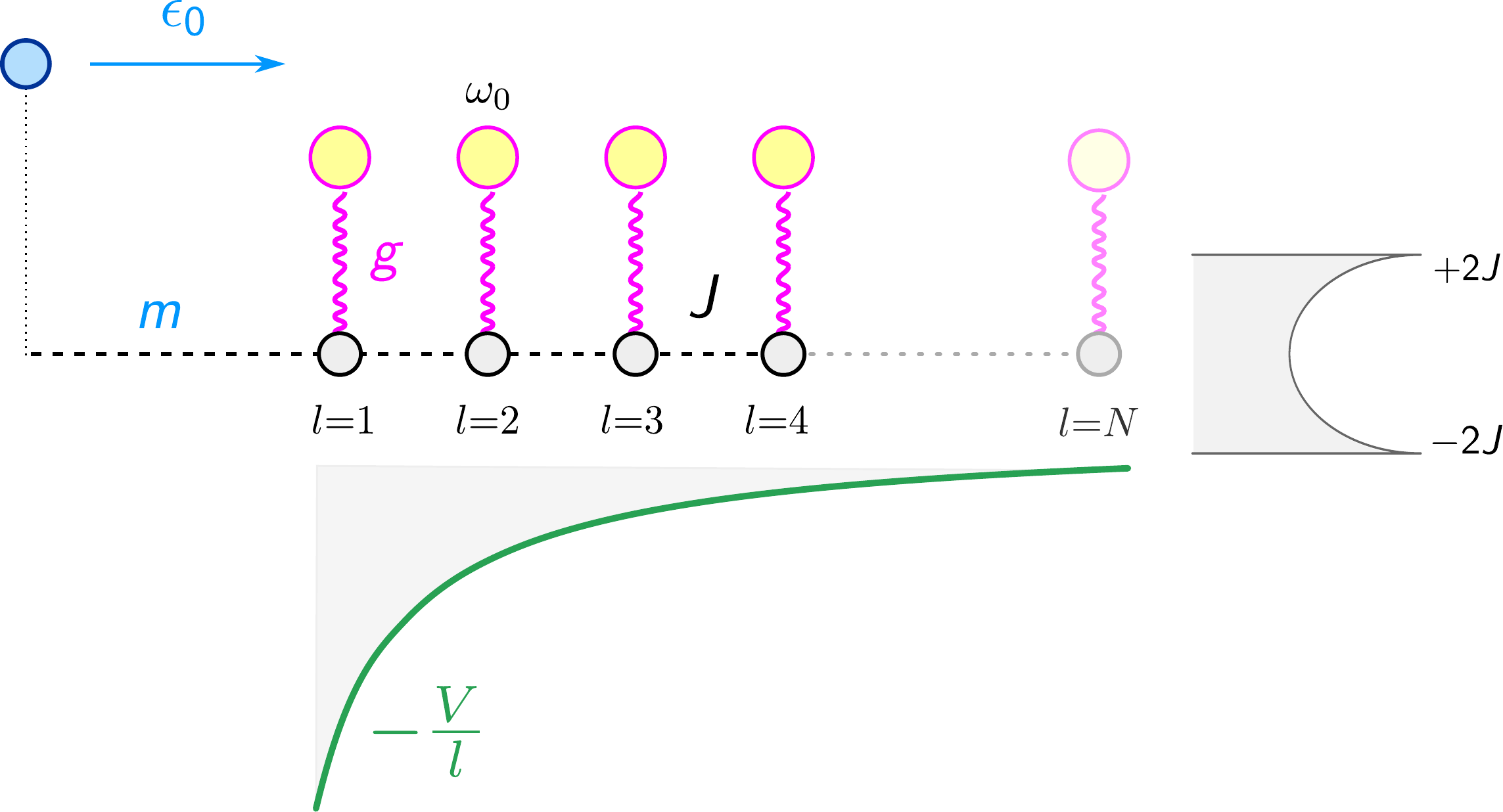}
  \caption{Schematic diagram of the electron-vibration chain model that it used the describe the
  dynamics at the donor-acceptor interface~Eq.~\eqref{eq:Ham}. 
  $\epsilon_0$ is the electronic energy after photoexcitation on the donor side, and the green curve represents
  the attractive Coulomb potential from the hole, that we suppose to be localized at site $l=0$. 
  The density of states
  of the non-interacting chain is shown on the right.}
  \label{fig:diag}
}

\paragraph{Model.} 
We use the following Hamiltonian to describe the quantum-mechanical exciton
charge transfer process in a molecular chain of length $N$ (depicted on 
Fig.~\ref{fig:diag}):
\begin{eqnarray}
\nn
\mc{H} &=& \omega_0 \sum_{l=1}^{N} a_l^\dg a_l^\pdg + 
g\sum_{l=1}^{N} c_l^\dg c_l^\pdg (a_l^\pdg  + a_l^\dg) 
- m (c_0^\dg c_1^\pdg + c_1^\dg c_0^\pdg) \\
&+& \epsilon_0 c_0^\dg c_0^\pdg 
- \sum_{l=1}^N \f{V}{l} c_l^\dg c_l^\pdg
- J\sum_{l=1}^{N-1} \Le(c_l^\dg c_{l+1}^\pdg + c_{l+1}^\dg
  c_l^\pdg \Ri)\!.
\label{eq:Ham}
\end{eqnarray}
Here $a_l^\dg$ and $c_l^\dg$ are respectively the phonon and electron creation
operators on site $l$, $\omega_0$ is the energy of the relevant molecular
vibration, $g$ is the electron-phonon coupling constant,
$m$ is the tunneling amplitude between the donor (site $l=0$) and the first site of the
tight binding chain describing the acceptor, $\epsilon_0$ is the energy of the 
incoming electron (related to the difference between the donor and acceptor band offset 
and the exciton binding energy), $V$ sets the typical Coulomb 
potential binding the electron-hole pair, and $J$ is the hopping amplitude within 
the chain. In the following paragraphs we shall use the dimensionless Huang-Rhys parameter 
$\alpha^2=(g/\omega_0)^2$ --- i.e. the ratio between the molecular reorganization energy $\lambda$ and
the vibrational energy $\omega_0$ ---
to characterize the strength of the electron-phonon interaction, as appropriate
in the non-adiabatic regime $\omega_0 \gtrsim J$ \cite{Feinberg,Capone}.

\begin{figure*}[!htb]
\includegraphics[width=1\textwidth,height=5.5cm]{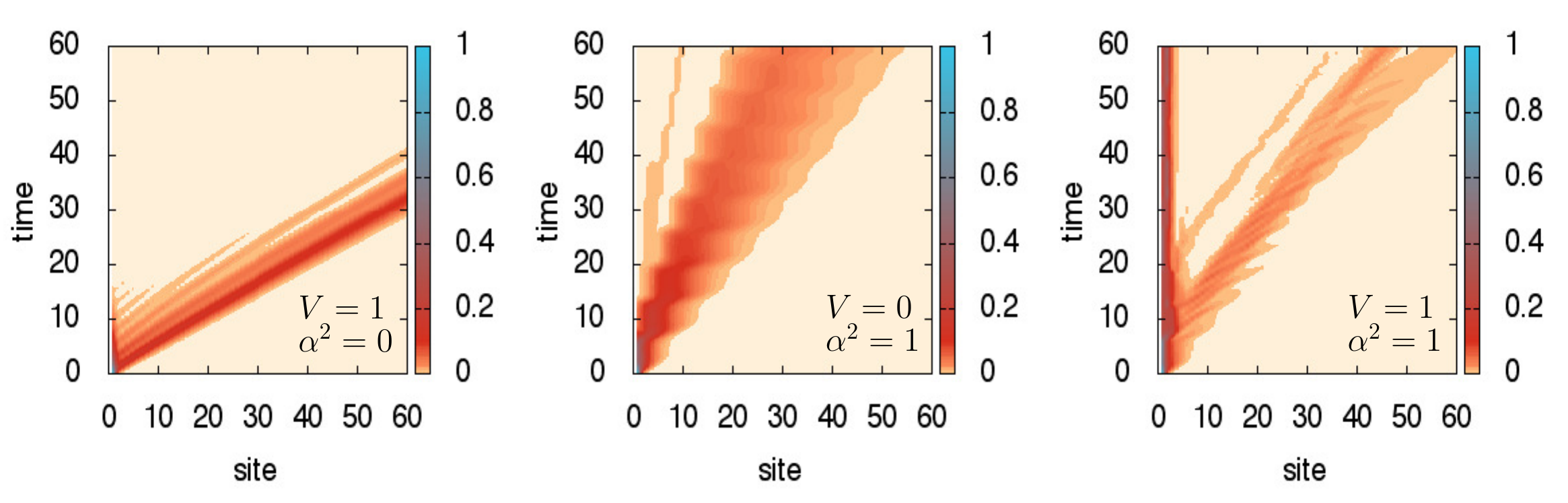}
\caption{(color online) Flow of electron density $|A_l(t)|^2$ at each site $l$ as a
function of time $t$ (in units of $1/J\simeq 20$fs), for an injection energy 
$\epsilon_0 = 0.1$ that is taken within the available band of delocalized states.
Three different dynamics are performed: left panel, Coulomb confinement only
($\alpha^2=0.0$, $V=1.0$); middle panel, electron-phonon interaction only
($\alpha^2=1.0$, $V=0.0$); right
panel, both interaction and confinement ($\alpha^2=1.0$, 
$V=1.0$). Only the latter case shows prominent localized states near the 
interface, that are however superposed with out-going wavepackets, giving 
a reduced but finite yield.}
\label{fig:maps}
\end{figure*}

We consider here simultaneously the combined effect of charge transfer
processes near the interface and the delocalized motion away from it,
in contrast to previous studies focusing either on the
former~\cite{Bittner,Tamura1} or the latter~\cite{Ratner,Wu,Troisi,Gelinas} issue.
First, we will not model the hole dynamics on the donor side, assuming that it is 
localized on site $l=0$.  
The effects of hole diffusion, that could be easily described by extending
Hamiltonian Eq.~(\ref{eq:Ham}), typically amounts to a
reduction of the effective Coulomb binding.
Second, a tight-binding model on a one-dimensional chain is used to describe a complex
bulk molecular heterostructure. The three-dimensional nature of the system and the 
realistic molecular orbitals are expected to lead to mainly quantitative changes 
in the bandstructure, which are currently the subject of theoretical
investigation~\cite{Gelinas,Bittner,Tamura2,Duchemin,TroisiPNAS,Smith2}. 
Third, we assume a single intramolecular mode of vibration, as vindicated
by the few high frequency discrete modes of 
fullerene molecules~\cite{Antropov,Faber,Canton} (in the narrow range $0.18-0.2$ eV) that account 
for the majority of the electron-phonon coupling.
The electron dynamics considered here concerns the fast electron
unbinding from the Coulomb well, occurring on the $10$ to $100$ femtosecond 
time scale~\cite{Falke,Song}.
We thus neglect the impact of weakly coupled low frequency phonon 
modes~\cite{Crespi,Antropov,Kolli,Smith}, 
as these will lead to final equilibration at longer relaxation times than 
relevant for the exciton dissociation process. 
Fourth, we take a pure Coulomb law for the electron-hole binding potential, but
any other type of interaction should not drastically affect our general results
(see Refs.~\cite{Duchemin,Tamura4} for a more microscopic approach).
We also choose to ignore disorder-induced localization effects, 
~\cite{Ratner,Smith,Troisi}, 
since a  localization mechanism is already provided by the interfacial binding
potential.
Finally, we study the dynamical evolution of the system at zero temperature, which 
is justified because both electronic and vibrational energy scales are much 
larger than $k_BT$ \cite{note-hop}.

\paragraph{Theoretical approach.} 
Model Eq.~(\ref{eq:Ham}) and variants have been the topic of several
recent theoretical studies
focusing either on detailed aspects of the exciton dissociation at the 
interface~\cite{Tamura1,Bittner}
or on dynamics of charge separation on the acceptor 
chain~\cite{Ratner,Gelinas,Wu,Smith}.
Our goal is to provide a joint quantum dynamical description for the electron
and vibrational modes on the chain --- whereas vibrational dynamics is often approximated by 
a classical evolution or by an effective decay rate on the electron motion --- and to explore the 
resulting microscopic phenomena in a broad range of relevant parameters.
The approach that we propose is the use of Dirac-Frenkel's time-dependent
variational principle~\cite{Kramer}, which has been applied with success to study ground state
and dynamical properties of a wide class of open quantum
systems~\cite{Ye,Bera,Zhao1,Zhao2}.
We will start with the following wavefunction Ansatz:
\begin{equation}
\label{eq:ansatz}
|\Psi(t)\big> = 
\sum_{l=0}^N A_l(t) c_l^\dg 
e^{f_l^\pdg\!(t) a_l^\dg - f_l^\star(t) a_l^\pdg} |0\big>.
\end{equation}
Here $A_l(t)$ corresponds to the local electron complex probability amplitude
at site $l$ and time $t$, while the complex parameter $f_l(t)$ describes a coherent 
state of the phonon mode $a_l$ ($|0\big>$ is the full fermionic and bosonic vacuum). 
We emphasize that the dynamics in the wavefunction~ Eq. (\ref{eq:ansatz}) is restricted to 
a single electron propagating along the whole chain, hence reducing drastically
the fermionic Hilbert space to the size $N+1$. However, the full electron-phonon 
problem is non-trivial, because for finite values of $\alpha^2$, several Fock states 
have to be considered per site, making the total number of states exponentially large.
An important rationale to vindicate Ansatz~ Eq. (\ref{eq:ansatz}) is that it 
encompasses several known limits. First, in the absence of
phonons ($\alpha^2=0$), the exact electron dynamics is obtained in
the presence of any binding potential (or even disorder). Second,
when charge hopping is turned off ($J=0$), local phonons experience
a displacement $f_l$ depending on whether a charge is present or not, an
effect that is exactly contained in our wavefunction.
A crucial aspect of the Ansatz is that it correctly captures 
the so-called non-adiabatic regime~\cite{Feinberg,Capone}, whereupon the phonon frequency
$\omega_0$ is not small compared to the tunneling energy, which 
is the case for the intra-molecular vibrations considered here.

We describe the time evolution of wave-function Eq. (\ref{eq:ansatz}) 
by constructing the following Lagrangian:
($\hbar=1$ in what follows):
\begin{eqnarray}
\mc{L} & = & \big<\Psi(t)|\frac{i}{2} \overleftrightarrow{\partial_t} 
- \mc{H}|\Psi(t)\big>\\
\nn
&=& \frac{i}{2} \sum_{l=0}^N |A_l|^2 
[ f_l^\star \dot{f}_l^\pdg - 
\dot{f}_l^\star f_l^\pdg ] + 
\frac{i}{2} \sum_{l=0}^N 
[ A_l^\star \dot{A}_l^\pdg-
\dot{A}_l^\star A_l^\pdg]\\
\nn
&&+ 
\sum_{l=1}^{N} |A_l|^2 \left[ \omega_0 |f_l|^2
+ g(f_l^\pdg+f_l^\star) + \epsilon_l 
\right] \\
&&- \sum_{l=1}^N J_l \left[ 
A_l^\star A^\pdg_{l+1} + 
A_{l+1}^\star A^\pdg_{l} \right]
e^{-|f_l|^2/2-|f_{l+1}|^2/2}
,
\end{eqnarray}
where we use the compact notation 
$\overleftrightarrow{\partial_t} = \overrightarrow{\partial}\!/\partial t
- \overleftarrow{\partial}\!/\partial t$,
$\dot{A}_l = \partial A_l /\partial t$,
$\dot{f}_l = \partial f_l /\partial t$,
$\epsilon_l = \epsilon_0 \delta_{l,0}
-\frac{V}{l}(1-\delta_{l,0})$,
and $J_l = m \delta_{l,0} + J (1-\delta_{l,0})$.
The full dynamical evolution can then be obtained from 
Hamilton-Jacobi equations:
\begin{eqnarray}
\frac{\mrm{d}}{\mrm{d}t} 
\frac{\partial \mc{L}}{\partial \dot{f}_i^\star} &=&
\frac{\partial \mc{L}}{\partial f_i^\star},\\
\frac{\mrm{d}}{\mrm{d}t} 
\frac{\partial \mc{L}}{\partial \dot{A}_i^\star} &=&
\frac{\partial \mc{L}}{\partial A_i^\star},
\end{eqnarray}
which read explicitly as:
\begin{eqnarray}
\nn
i \dot{f}_l &=& 
\frac{J_l f_l}{A_l} e^{-|f_l|^2/2}
\left[A_{l+1}e^{-|f_{l+1}|^2/2}+
A_{l-1}e^{-|f_{l-1}|^2/2}\right]\\
\label{eq:fdot}
&&+ \omega_0 f_l + g \equiv F_l, \\
\nn
i \dot{A}_l & = & -\frac{A_l}{2} \left[f_l F_l^\star+f_l^\star F_l\right]
+ A_l\left[\omega_0 |f_l|^2
+ g(f_l^\pdg+f_l^\star) + \epsilon_l\right]\\
\label{eq:Adot}
&-&
J_l e^{-|f_l|^2/2}
\left[ A_{l+1} e^{-|f_{l+1}|^2/2}
+ A_{l-1} e^{-|f_{l-1}|^2/2}\right]\!,
\end{eqnarray}
where $f_0=0$ is enforced since we
do not include phonons on the donor site. 
Note that the local displacement of the phonon
$X_l(t) = \big<\Psi(t)|a_l^\dg+a_l^\pdg|\Psi(t)\big> = |A_l(t)|^2 
[f_l^\star(t)+f_l^\pdg(t)]$ in the present notation. In the limit of
isolated molecular sites ($J=0$), the electron density is time-independent, and
we find the physically correct vibrational dynamics:
$\dot{P_l} = \omega_0 X_l + 2 g |A_l|^2$, with $P_l$ the conjugate
momentum to $X_l$.
The closed set of dynamical equations~(\ref{eq:fdot})-(\ref{eq:Adot}) 
can be efficiently and accurately solved using a fourth-order Runge-Kutta
algorithm. Typically, a time step of $\delta t=0.01$
in the natural units ensures good convergence. Of particular importance to
check the quality of the numerical solution is to monitor the norm $N=
\big<\Psi(t)|\Psi(t)\big>$ and the average energy 
$E=\big<\Psi(t)| \mc{H}|\Psi(t)\big>$, and verify that they do not
evolve with time, as these are exactly conserved quantities during the 
dynamics (as can be easily proven from the previous equations).

\paragraph{Results.}
Our initial state consists of an electron injected on the donor site 
$l=0$ with excess energy $\epsilon_0$,
without any lattice distorsion on the acceptor side, namely  
$|\Psi(t=0)\big> = c_0^\dg |0\big>$. We set the basic energy unit to be
the hopping parameter $J=1$.
We take the donor to acceptor
tunneling rate $m=0.5$, having checked that our results do not depend
much on this quantity as long as $m\lesssim0.5$. We also fix the vibration frequency
$\omega_0=1$, that we take in the realistic experimental range, but again the
precise value is not very important as long as one remains in the 
non-adiabatic regime. We leave free the other three parameters i.e. the injection energy
$\epsilon_0$, the electron-phonon vibronic coupling $\alpha^2$ and the Coulomb strength $V$.

We start by investigating the time-evolution of the  wavefunction along
the chain as time evolves. On Fig.~\ref{fig:maps} we consider 
the dynamics of the electron density
$\big<\Psi(t)|c_l^\dg c_l^\pdg|\Psi(t)\big>= |A_l(t)|^2$.
The leftmost panel considers the sole effect of Coulomb
confinement ($\alpha^2=0.0$, $V=1.0$), and our results here are numerically exact, by
construction of Ansatz~ Eq. (\ref{eq:ansatz}). We find that the main contribution is in
the form of pulses of ballistic wavepackets emitted from the interface. This
bevavior is readily explained by the undamped oscillatory motion of the electron
in the confinement zone, which leads to periodic emission of electron probability 
due to resonant scattering by the external potential.
Clearly, the probability weight near the interface is progressively decaying with time, and
the particle is fully delocalized, a picture often proposed in the literature
\cite{Ratner,Gelinas,Troisi},
that however drastically overestimates the energy transfer efficiency of
realistic devices.
When the electron-phonon interaction is switched on, but without binding potential 
(middle panel, with $\alpha^2=1.0$, $V=0.0$), two striking effects arise 
due to the dressing of electronic states by phonons (polaron formation).
First, for most values of the injection energy (here shown $\epsilon_0=0.1$), the beam 
is dominated by a fast ballistic component, albeit with a strong renormalization and 
dispersion of velocities. 
Second, temporal oscillations of period $2\pi/\omega_0$ of the electron transfer probability
appear, clearly associated to vibronic coupling.

The right panel in Fig.~\ref{fig:maps} shows the most interesting case where both interaction and 
confinement are taken into account ($\alpha^2=1.0$, $V=1.0$). 
While the presence of molecular vibrations or Coulomb binding 
taken alone do not prevent the migration of electrons to the electrodes,
their combined effect provides strong interfacial relaxation, leading to prominent 
localized states near the interface. 
The observed collapse of the exciton wavefunction in proximity 
of the interface drastically affects the transmission, which yet remains
finite, as is seen from the existence of a ballistic beam of electrons with
reduced amplitude (compared to the left and middle panels). 
As seen in Fig.~\ref{fig:maps},  
charge transfer begins on a time scale of a few tens of fs (the time unit is 
$1/J\simeq 20$fs, where we have assumed an effective value $J= 0.2$eV
corresponding to the typical electronic bandwith $W\simeq 0.8$eV of fullerene
materials). The coexisting localized and delocalized components both 
show coherent oscillations caused by the vibronic coupling, 
and become fully spatially separated after roughly 200fs.

\fig{
  \includegraphics[width=0.9\columnwidth]{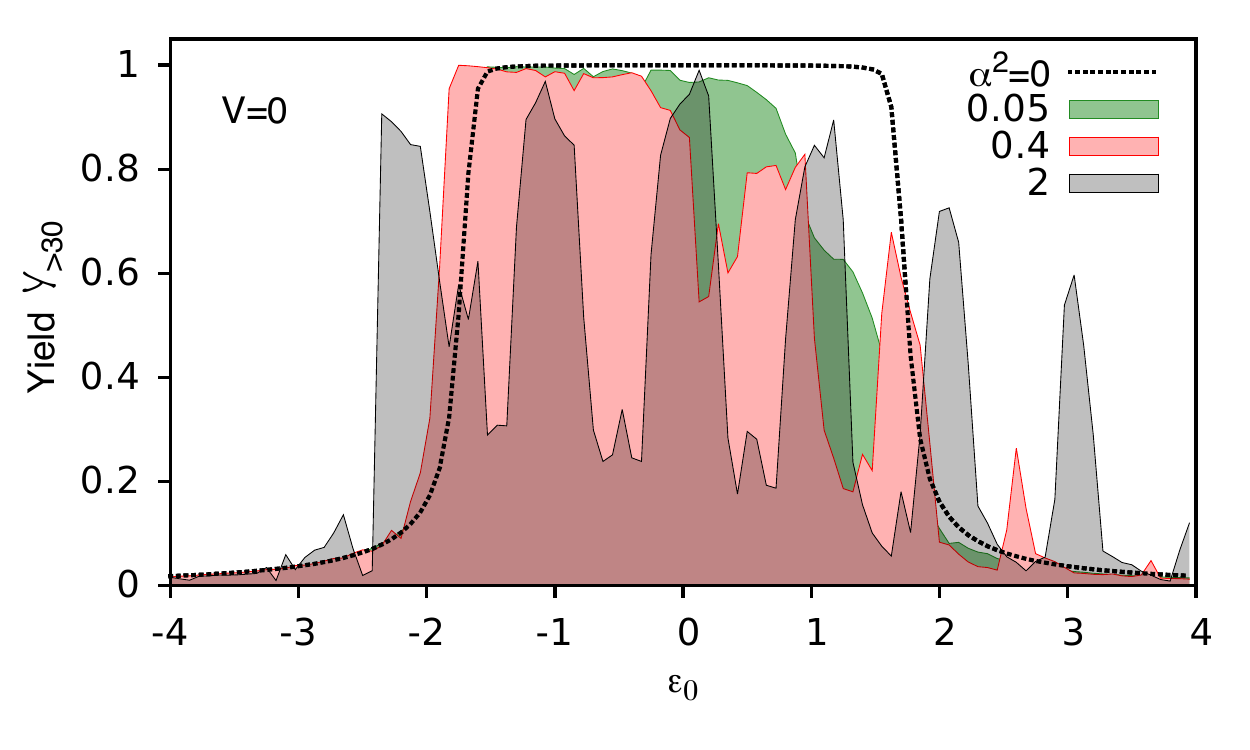}
  \includegraphics[width=0.9\columnwidth]{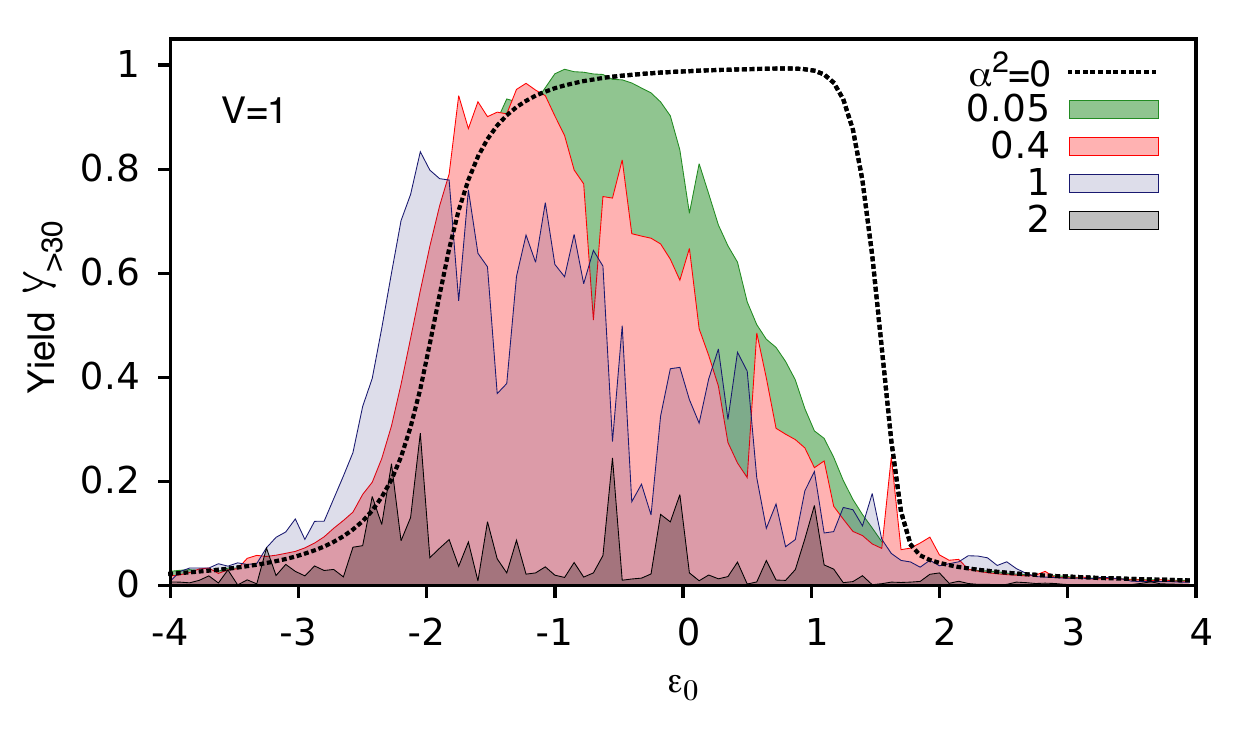}
  \caption{(color online) Yield $\Yield$ as a function of incoming 
energy $\epsilon_0$, for various values of electron-phonon coupling constant 
(the non-interacting case  $\alpha^2=0$ is shown as dashed line).
The upper panel is the case without electron-hole binding ($V=0$), 
the lower one has $V=1$. The combined effect of Coulomb binding and 
polaronic dressing of the carriers leads to a strong suppression of the yield. 
}
\label{fig:yield}}

We  now  quantify more precisely the degree of charge delocalization,
by defining a simple measure of the efficiency for energy transfer in the device.
To this aim we consider the site-dependent yield as the total electron
probability collected in the long-time limit beyond the site $l=L$ ~\cite{note}:
\begin{equation}
\mc{Y}_{>L} = \lim\limits_{t\to\infty} \sum_{l>L}|A_l(t)|^2.
\end{equation}
In practice, we choose $L=30$, which is far enough from the interface and
avoids the influence of the interfacial bound states.
In addition, $L$ extends beyond the room temperature Coulomb thermal
length defined by $L_T=V/(k_B T)\simeq12$ for a typical well of depth 0.3 eV,
ensuring that charge can be considered as free beyond this range.
We also found (for the parameters used here) that a maximal time $t=200$ is sufficient to
obtain converged values for $\mc{Y}_{>L}$. One must in addition take a
chain of about $N=200$ sites in order to prevent the wavepacket from bouncing
off the boundary, the portion of the chain $30<l<200$ thus playing the role
of a perfect sink.

We start by considering the case without interfacial confinement ($V=0$) in the
upper panel of Fig.~\ref{fig:yield}, scanning for all possible values of
the injection energy $\epsilon_0$ and changing the strength of the electron-phonon 
interaction $\alpha^2$
(we show in particular $\alpha^2\simeq0.4$, the value corresponding to 
fullerene molecules~\cite{Faber}).
In the non-interacting case, shown for reference, the yield is essentially unity within  the bare 
acceptor bandwidth $-2J \lesssim \epsilon_0 \lesssim 2 J$ \cite{Ratner}.
A moderate reduction in the yield is observed 
already at weak electron-phonon coupling,  $\alpha^2 = 0.05$, affecting 
the high energy side of the band  
(the lower band edge is unaffected, as expected 
from the existence of long-lived delocalized states at
the band bottom~\cite{Ciuchi2}). 
Upon increasing $\alpha^2$, the dressing of 
carriers by large molecular deformations leads to the emergence of multiple 
sub-bands~\cite{Ciuchi2} separated in energy by $\omega_0$,
which reflect the polaronic structure of the acceptor band.
The yield is consequently strongly modulated and overall 
redistributed over a wider injection energy window. 
We emphasize that for any value of $\alpha^2$, however large, one can find 
energy bands where the yield is close to one, showing that polarons behave as 
nearly ballistic charge carriers in a homogeneous system (cf. Fig. 2, middle panel).

Bringing in the effect of electron-hole binding (with $V=1$) is shown in the lower 
panel of Fig.~\ref{fig:yield}. While periodic resonances from polaronic dressing are still visible,  
the global reduction of the yield is clearly
dramatic, as could be anticipated from the right panel of
Fig.~\ref{fig:maps}. A sharp drop of the  overall efficiency occurs  with increasing $\alpha^2$, not observed 
in the absence of Coulomb binding.
Comparison of the two panels in Fig.~\ref{fig:yield} shows that the effects
of phonon relaxation and interfacial confinement do not simply add up, and
that the Coulomb potential has a dramatic interplay with polaronic formation:
relaxation processes brought in by the molecular vibrations
cause a collapse of electronic states towards localized levels near the interface, 
strongly suppressing the charge delocalization.

The overall photogeneration efficiency of a given device will result from 
both the efficiency of photon absorption in the required energy 
range and the ability of excitons to dissociate at the interface.
Having addressed the latter process via model Eq.~(\ref{eq:Ham}), we can conclude 
that the device performances are strongly suppressed in the presence of strong 
electron-phonon coupling, indicating that organic molecules with low reorganization 
energies should be used preferentially.
Also, it is apparent from Fig.~\ref{fig:yield} that high energy electronic states 
are generally  more prone to relaxation, so that 
the  injection of ``hotter'' carriers should not lead to higher device 
performances in the presence of vibronic coupling.

In conclusion, we have provided microscopic evidence that the efficiency 
for energy transfer of electrons across organic interfaces is 
subtly controlled by an interplay of electrostatic confinement
and coherent coupling of electrons to high energy quantized vibrational 
modes.
Our findings rationalize the oscillatory behavior of transferred 
charges due to molecular vibrations, as recently observed in 
Refs.~\cite{Falke,Song}. They also reconcile the conflicting views often 
presented in the literature, arguing in favor of either prominently localized
or fully delocalized states~\cite{Bakulin,Gelinas}.
This physics may also be of general relevance to a wider class of
light harvesting systems than organic solar cells, designed from 
inorganic~\cite{rozbicki} or biological~\cite{scholes2011} elements, where 
charge separation occurs also in presence of strong electron-phonon coupling.

\acknowledgments
We thank X. Blase, A. Salleo and K. Vandewal for useful discussions.


\begin{thebibliography}{99}
\bibitem{note}
This quantity is related, but not equal, to the internal quantum efficiency 
commonly extracted from experiments: one can not directly relate the injection energy 
$\epsilon_0$ in Eq. (1) to the incoming photon frequency, because 
a continuum of possible initial electronic states is generated in 
the HOMO-LUMO bands upon absorption of light.
\end{thebibliography}
\end{document}